\documentclass[%
 reprint,
superscriptaddress,
 amsmath,amssymb,
 aps,
prb,
]{revtex4-1}

\usepackage{graphicx}
\usepackage{dcolumn}
\usepackage{bm}
\usepackage{amsmath,amssymb}
\usepackage{graphicx}
\usepackage{color}

\begin{document}
\preprint{APS/123-QED}
\title{Fabrication of porous microrings via laser printing and ion-beam post-etching}

\author{S. Syubaev}
\affiliation{School of Natural Sciences, Far Eastern Federal University, 6 Sukhanova Str., 690041 Vladivostok, Russia}
\affiliation{Institute for Automation and Control Processes FEB RAS, 5 Radio Str., 690041 Vladivostok, Russia}

\author{A. Nepomnyashchiy}
\affiliation{Institute for Automation and Control Processes FEB RAS, 5 Radio Str., 690041 Vladivostok, Russia}

\author{E. Mitsai}
\affiliation{Institute for Automation and Control Processes FEB RAS, 5 Radio Str., 690041 Vladivostok, Russia}

\author{E. Pustovalov}
\affiliation{School of Natural Sciences, Far Eastern Federal University, 6 Sukhanova Str., 690041 Vladivostok, Russia}

\author{O. Vitrik}
\affiliation{School of Natural Sciences, Far Eastern Federal University, 6 Sukhanova Str., 690041 Vladivostok, Russia}
\affiliation{Institute for Automation and Control Processes FEB RAS, 5 Radio Str., 690041 Vladivostok, Russia}

\author{S. Kudryashov}
\affiliation{Lebedev Physical Institute, Russian Academy of Sciences, 53 Leninsky prospect, Moscow 119991, Russia}
\affiliation{ITMO University, 49 Kronverksky prospect, St. Petersburg 197101, Russia}

\author{A. Kuchmizhak}
\email{alex.iacp.dvo@mail.ru}
\affiliation{School of Natural Sciences, Far Eastern Federal University, 6 Sukhanova Str., 690041 Vladivostok, Russia}
\affiliation{Institute for Automation and Control Processes FEB RAS, 5 Radio Str., 690041 Vladivostok, Russia}

\begin{abstract}
Pulsed-laser dry printing of noble-metal microrings with a tunable internal porous structure, which can be revealed via an ion-beam etching post-procedure, was demonstrated. Abundance and average size of the pores inside the microrings were shown to be tuned in a wide range by varying incident pulse energy and a nitrogen doping level controlled in the process of magnetron deposition of the gold film in the appropriate gaseous environment. The fabricated porous microrings were shown to provide many-fold near-field enhancement of incident electromagnetic fields, which was confirmed by mapping of the characteristic Raman band of a nanometer-thick covering layer of Rhodamine 6G dye molecules and supporting finite-difference time-domain calculations. The proposed laser-printing/ion-beam etching approach is demonstrated to be a unique tool aimed at designing and fabricating multifunctional plasmonic structures and metasurfaces for spectroscopic bioidentification based on surface-enhanced infrared absorption, Raman scattering and photoluminescence detection schemes.

\vspace{0.1 cm}
\end{abstract}

\maketitle
Surface-enhanced Raman scattering (SERS) is an ultra-sensitive non-invasive spectroscopic technique based on a label-free identification of different molecules placed in the vicinity of plasmonic-active nanostructured metallic substrates ~\cite{Jahn16,Anker08,Chung11,Kuchmizhak2016}. Intensity of the characteristic Raman signal defining the specific vibrational signatures of individual molecules is usually very weak. However, this signal can be significantly increased near nanotextured surfaces or nanostructures generating localized, strongly enhanced plasmon-mediated electromagnetic fields. Since the first observation of SERS signal from single molecule \cite{Nie97}, multiple attempts were undertaken to increase the efficiency of SERS-active nanotextured substrates in terms of achieved maximal enhancement factor inside a single ``hot spot'' as well as number (density) of ``hot spots'' per individual nanostructure~\cite{Wei16,Liu16,JWi12,Yan16,Zeng15,Yashchenok12}.

To address both issues, variety of nanotextured structures, predominantly having large surface-to-volume ratio and generating dense hot spots (tipped structures, nanostructures with inner porosity, intra-gap structures or self-assembly superstructures, etc.) were fabricated and tested as versatile SERS substrates\cite{Hamon16,Nie09,Yang14,Vidal15,Zhang2014,Zhu15,Malgras15,Zhou15,Jeon13,Jones11,Zhang2k14,Sun13}. Specifically, porous materials, nanostructures and nanoparticles, routinely reaching uniform SERS enhancement sufficient to overcome single-molecule detection limit independently on excitation/detection conditions, are of growing interest \cite{Sahoo14,Ye16,Liu11}. Several papers reported fabrication of such sponge-like structures, using dealloying of the two-component template via its dissolving in a corrosive environment \cite{Forty79,Wang12}. Meanwhile, to design sensitive elements for advanced biosensors, along with desirable porosity, it is also important to control the overall size and shape of such porous templates as well as to arrange them into well-ordered arrays at specific point on a substrate with micrometer-scale lateral accuracy. Despite the latter issue can be resolved by using well-established but rather time-consuming non-scalable and expensive electron- or ion-beam lithography techniques \cite{Ahmadi13,Li01}, contamination of nanotextures during wet etching procedure as well as complete dealloying of the fabricated nanostructures present issues, which are still to be resolved.

Pure chemical techniques as seed-mediated growth, chemical vapor deposition or electrodeposition are suitable only for cheap inexpensive nanofabrication of disordered porous nanotextures and nanoparticles \cite{Zhang14,Tsai11,Schubert15}, providing high controllability of geometric shapes and porosity, but requiring, however, additional pre-processing fabrication steps to arrange the isolated elements into their ordered arrays. Rapid laser-assisted thermal annealing and post-dealloying enable higher performance in comparison to the above mentioned approaches, requiring, on the other hand, additional steps in the processing chain, and, more importantly, often cause undesirable decrease of pore density \cite{Arnob14,Zeng14}. In this way, high-performing, easy-to-implement, ``dry" and ``green" (without hazardous chemicals) technology for fabricating porous plasmonic nanostructures is still missing.

In this Letter, an ion-beam assisted, liquid-free nanosecond (ns) laser printing of isolated plasmonic ring-shaped nanostructures with pronounced porosity is demonstrated. Besides the recently reported possibility to control and tune over the main geometric parameters of the laser-printed rings, such as diameter, wall thickness and height \cite{Kuchmizhak16}, herein we demonstrate that an additional degree of tunability over the inner structure of the fabricated microrings can be achieved via utilization of a nitrogen-doped Au film. Variation of N$_{2}$ concentration in a buffer gas during the process of magnetron deposition of such Au film was demonstrated to pre-determine size and density of resulting nanopores, appearing as a result of boiling inside the molten rim produced upon single-shot ns-laser film ablation. The contours of the microring as well as the inner pores can be further unveiled in the process of ion-beam post-polishing, yielding in isolated rough plasmonic nanostructures with pronounced porosity. The fabricated porous microrings were shown to produce many-fold near-field enhancement of incident electromagnetic fields, which was confirmed via mapping of the characteristic Raman bands of a self-organized monolayer of the Rhodamine 6G (R6G) dye molecules and supporting finite-difference time-domain (FDTD) calculations.

\begin{figure}
\includegraphics[width=0.9\columnwidth]{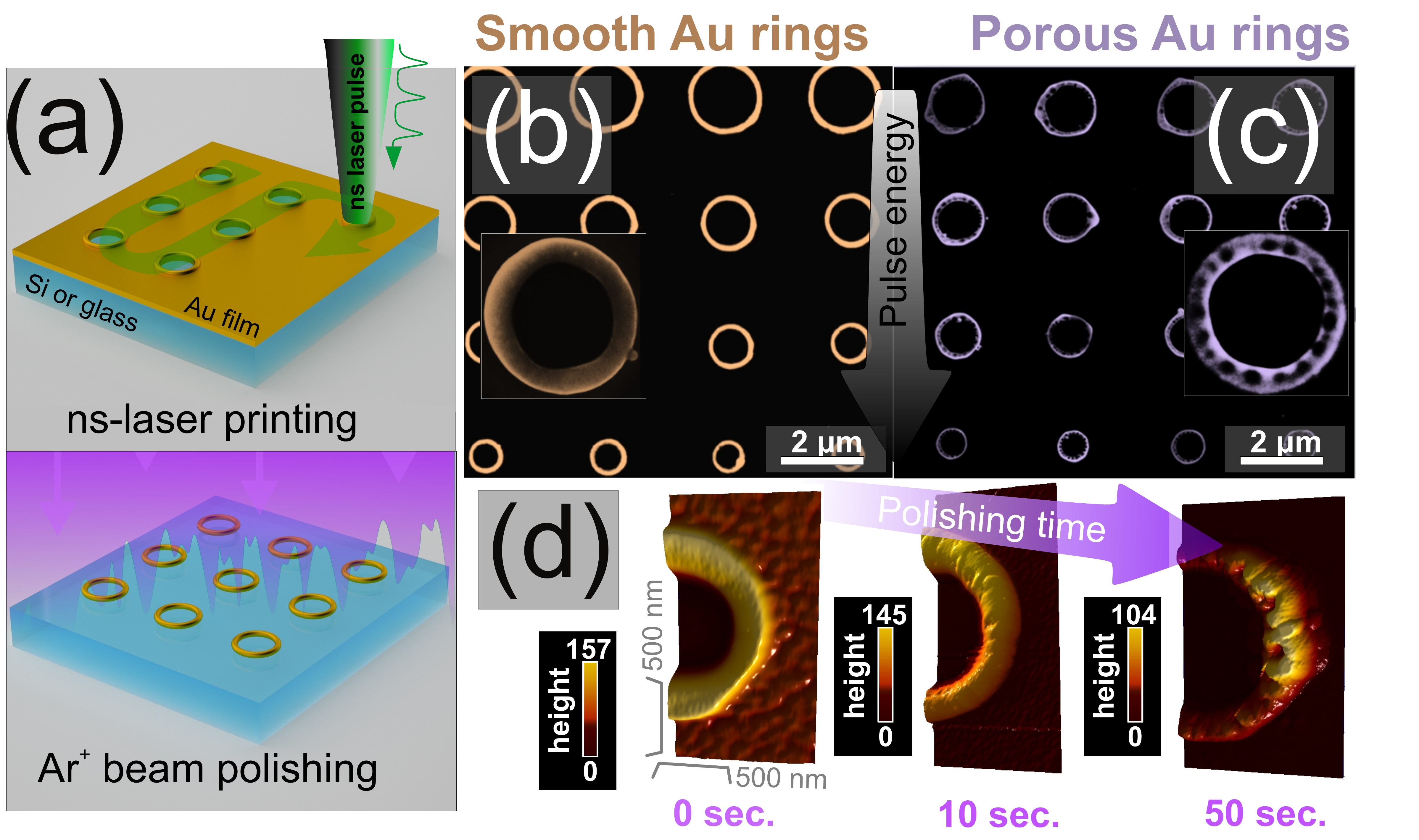}
\caption{(a) Schematic of the two-step fabrication procedure for printing of the isolated porous plasmonic rings; (b,c) Normal-view false-color SEM images of the smooth and porous rings fabricated using ion-beam assisted ns-laser printing (pulse energy decreases from upper to bottom rows); Insets on both images show magnified SEM images of isolated smooth and porous 1.5-$\mu m$ diameter Au microrings. (d) Series of AFM images showing the evolution of the roughness of the microring walls during the Ar$^{+}$-ion beam polishing. The nanopores initially located inside the microring walls are unveiling via the polishing post-procedure. Similar procedure performed to process the smooth microrings results in negligible changes of the initial roughness of the walls.
}
\end{figure}

The proposed procedure for fabricating the isolated porous rings includes two consecutive steps schematically illustrated in Fig.1(a). First, a 100-nm thick glass- or Si-supported Au film is irradiated with single second-harmonic (532 nm), 7-ns pulses delivered by a Nd:YAG laser system (Brio, Quantel) focused with a dry objective lens of numerical aperture NA=0.6. In all our experiments, we used the same focusing conditions, while the only  incident pulse energy was varied by means of an adjustable attenuator. In a certain range of incident pulse energies \cite{Kuchmizhak16}, such single-pulse ns-laser ablation of the metal film leaves a through hole surrounded by a smooth resolidified rim typically 2.5-3 times thicker, comparing to the initial thickness of the deposited Au film. The second step involves removal of the unprocessed parts of the film using a liquid-free etching with an accelerated argon-ion beam (IM4000, Hitachi) at fixed values of acceleration voltage of 3kV, discharge current of 105 $\mu$A and gas flow of 0.1 cm$^3$/min. These parameter set provides relatively slow rate of 1 nm/s to avoid film melting. The ion-beam etching procedure was preliminary calibrated in terms of etching rate using as-deposited gold films of different thickness (for additional details, see Ref.\cite{Kuchmizhak16}). The removal of the residual, initially deposited 100-nm thick Au film leaves the isolated microrings on the substrate surface. In our previous study, main geometric parameters of the microrings produced with the developed two-step procedure were shown to be tuned by varying the thickness of the as-deposited film and the incident pulse energy \cite{Kuchmizhak16,Kuchmizhak15}. In the present study, we demonstrate that an additional degree of tunability of the internal structure of the produced microrings can be provided by minor variation of the chemical composition of the sputtered metal film via its controlled magnetron deposition in appropriate gaseous environments. To deposit the Au film on a glass substrate a commercial magnetron sputter was used (Quorum Technologies)\cite{note2}. Typically, argon is used as a discharge gas for such magnetron sputtering. Then, ns-laser printing followed by the subsequent ion-beam polishing usually results in formation of smooth microrings with a relatively small amount of defects demonstrated by their detailed SEM inspection (see Fig.1(b)).

\begin{figure}
\includegraphics[width=0.82\columnwidth]{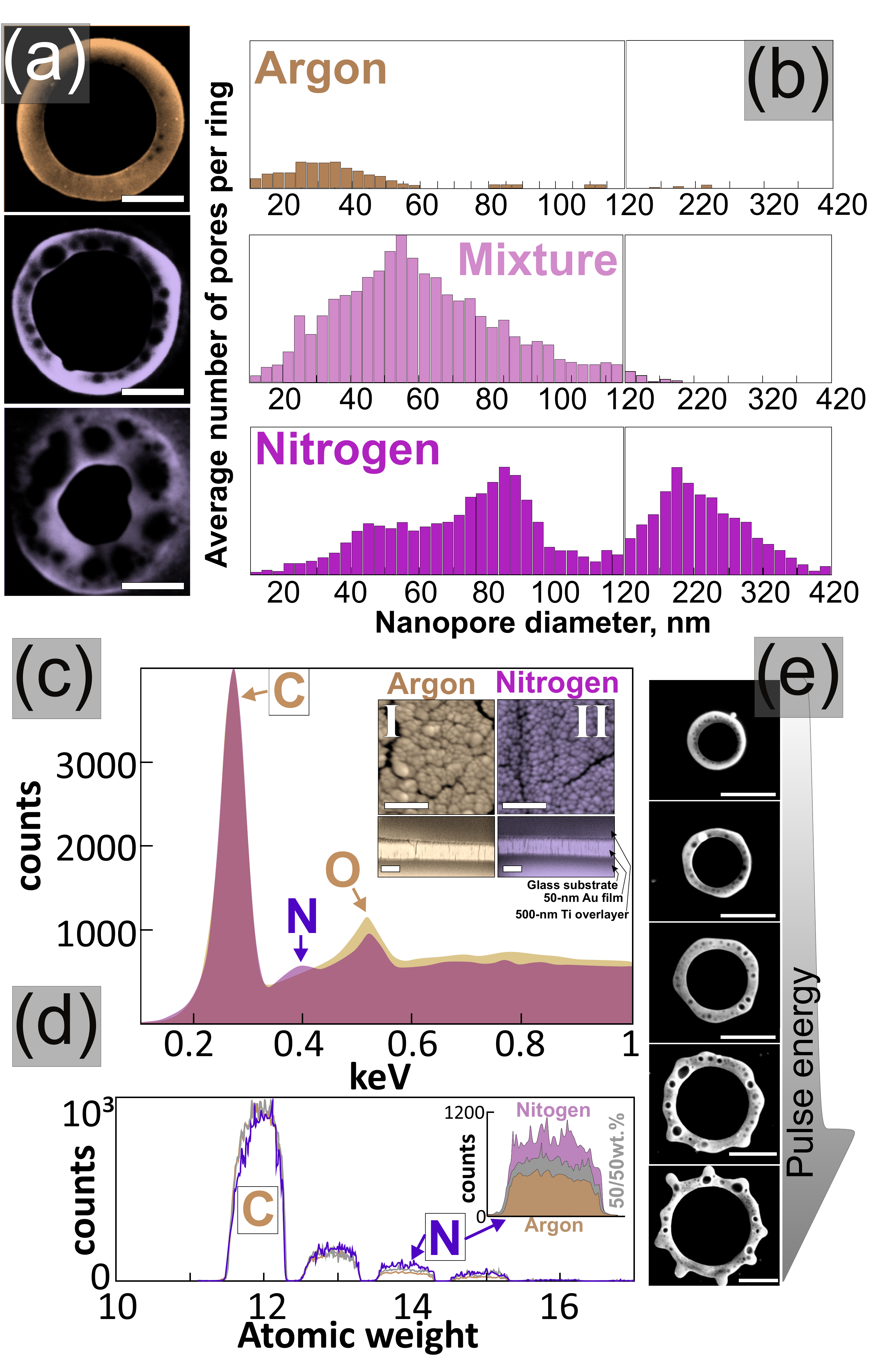}
\caption{(a) Series of normal-view false-color SEM images of microrings produced on the 100-nm thick Au films deposited in the argon (top), argon-nitrogen (50/50 wt.\%) gas mixture (middle) and nitrogen gas atmospheres (bottom) via magnetron sputtering. All three microrings were produced at the same pulse energy E=17 nJ. The scale bars are 500 nm. (b) Averaged distribution of nanopore diameters inside the fabricated microrings. The statistics for each film type was accumulated over 50 microrings of the same size produced under similar experimental conditions; (c) EDX spectra measured from the Au films deposited in the argon (brown) and nitrogen (purple) gaseous ambients. Grain structure as well as the multiple through cracks of both films are revealed via SEM inspection of the film surfaces (top view) and FIB cross-section cuts (bottom row) of the films deposited in the argon and nitrogen ambients. These images are presented as the insets I and II. (d) SIMS spectra of the films deposited in the argon (brown), argon-nitrogen gas mixture (gray) and nitrogen (purple) ambients. The magnified view of the nitrogen peak is given in the inset.
(e) A series of normal-view SEM images of the microrings produced at the increasing pulse energy on the surface of the 100-nm thick Au film sputtered in the argon-nitrogen gas mixture. The scale bar is 1 $\mu$m.}
\end{figure}

Surprisingly, detailed imaging of the structures produced with the single-shot ns-pulse ablative patterning of a Au film sputtered in a mixture of the argon and nitrogen (50/50 wt.\%) reveals dense pores inside the fabricated microrings (Fig.1(c)). The nanosized (appox. 60-nm diameter) spherical-shape pores initially located inside the resolidified metal rim can be further visualized through the ion-polishing post-procedure, as indicated by the series of the AFM images (Fig.1(d)). This series presents the several consecutive polishing steps from the as-irradiated film with the smooth walls to the complete removal of the non-treated regions of the film, revealing the micro-sized contours of the rings as well as their inner nanopores. Moreover, even more pronounced porosity with significantly larger averaged pore size was found for such structures produced on the surface of the Au film deposited in the nitrogen atmosphere (Fig.2(a)). The statistical study of the distribution of the pore size inside the microrings produced at the same pulse energy indicates the clear correlation between the nitrogen concentration in the deposition chamber and the average nanopore diameter in the produced microrings (Fig.2(a)).

\begin{figure}
\includegraphics[width=0.83\columnwidth]{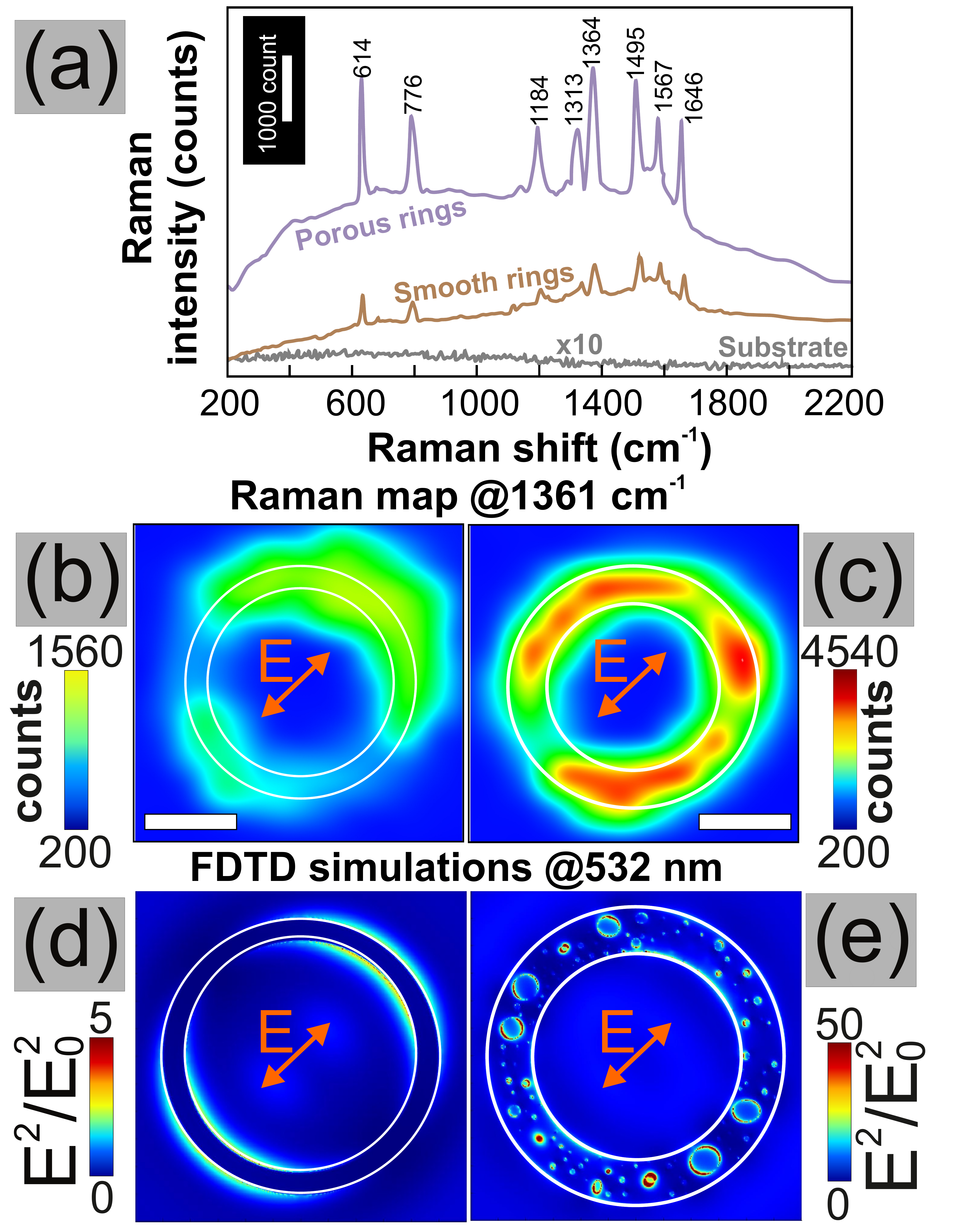}
\caption{Biosensing performance of smooth and porous microrings presented as R6G Raman spectra measured on the smooth (brown curve) and porous microrings (purple curve) as well as on the glass substrate after complete removal of the Au film via the ion polishing post-procedure. Each spectrum was averaged over 100 identical single-point spectra measured at similar experimental conditions; (b,c) SERS maps measured for the 1361-cm$^{-1}$ Raman band of the R6G molecules nanolayer covering the smooth (b) and porous microrings (c). These maps were found to provide SERS enhancement according to the characteristic E-field amplitude distribution near the smooth (d) and porous microrings (e) calculated for the normally-incident, linearly-polarized 532-nm laser radiation. The orange arrows indicate the polarization direction. The white circles show the actual dimensions of the microrings. The scale bar is 500 nm everywhere}
\end{figure}

Interestingly, detailed high-resolution SEM and AFM inspection of the untreated films sputtered in various gas environments indicates negligibly small variation of the film morphology (roughness, average grain size, etc.), comparing to the film deposited in the argon atmosphere\cite{Note1}. Specifically, for all these tested as-deposited films this analysis indicates the appearance of the through cracks, apparently owing to low-vacuum sputtering conditions ($\approx$10$^{-2}$ Torr) and relatively fast sputtering rate of 1 nm/s (see insets in Fig.2(c)). The remelting of such defects on the nanosecond-laser exposure timescale can potentially explain such appearance of the small amount of nanosized pores inside the rings produced in the ``Ar-deposited" films (Fig.2(a), top). However, considering the similarity of the morphologies of the film deposited in the various gaseous environments, such explanation can not be used for other films, demonstrating pronounced porosity and significantly larger average diameter of pores (Fig.2(a), middle and bottom\cite{note3}. In this way, subsurface boiling around such nitrogen-doped cracks and N$_2$-reach areas of the film appears to be the key mechanism responsible for the formation of the densely-packed nanopores inside the molten rim\cite{Kuchmizhak14}. As known, gold can interact with nitrogen, producing gold-nitride phase Au$_{x}$N, upon Au target bombardment with accelerated nitrogen ions \cite{Siller05,Brieva09}. Also, Au$_{x}$N films were fabricated, using ablative pulsed laser deposition in a N$_2$ containing atmosphere\cite{Caricato07}. Our energy-dispersive X-ray (EDX, ThermoDry) spectroscopic analysis performed at the 5-kV e-beam acceleration voltage under its oblique (47$^{\circ}$ to the sample normal) incidence shows the detectable nitrogen peak for the Au films sputtered in the nitrogen atmosphere (Fig.2(c)). Additionally, we have probed the same films by secondary-ion mass spectrometry (SIMS, Hyden Analytical EQS1000 installed in the Zeiss CrossBeam 1540 apparatus), utilizing a Ga$^{+}$ beam at 30 kV and 200 pA for sampling of 200x200 $\mu$m$^{2}$ areas of each film. For each sample, we averaged the obtained mass spectra over the depth profile. As seen, such comparative SIMS study reveals higher signals of nitrogen atoms (see Fig.2(d) and inset therein as well as Ref.\cite{note4}) in the films deposited in the nitrogen-containing ambients, while the carbon level for all films remains at the constant level. More generally, nitrogen-to-gold peak ratios averaged over the film depth are 0.23, 0.15, 0.14 for the films deposited in the nitrogen, nitrogen-argon mixture and argon ambients. Despite both methods indicate the increased amount of the nitrogen inside the metal films sputtered in the appropriate atmosphere (Fig.2(c,d)), the X-ray photoelectron spectroscopy should be undertaken during this ongoing research to prove the chemical bounding of nitrogen to gold.

Additionally, since the boiling of the nitrogen-rich areas of the Au film appears to be a temperature-driven process, the size of the nanopores also should demonstrate similar tendency. This is clearly illustrated by a series of SEM images of the microrings printed at the increasing pulse energy on the N$_2$-doped Au film (Fig. 2(e)), where the evident increase of the nanopore diameter can be identified. Specifically, in the certain range of the incident pulse energies E, the molten rim becomes hydrodynamically unstable undergoing the periodical modulation of its height (``crowning"\cite{Kulchin14,Qi16,Nakata09}) governed by the Rayleigh-Plateau hydrodynamic instability.
\indent The printed microrings, with their well-controlled micrometer-scale diameter and pronounced nanotexture governed by the unveiled multiple circumferentially-spaced pores, are expected to provide strong near-field enhancement of incident electromagnetic waves in visible and IR spectra ranges, making them promising for chemo- and biosensing applications. In this respect, the porous rings arranged into the ordered array can be considered as multifunctional substrates, providing enhancement of both photoluminescence and SERS signals from adsorbed analyte molecules commonly pumped in the optical range, while giving also possibility to utilize near-to-mid infrared wavelengths to excite main dipolar localized plasmon resonances of the microrings.

Without loss of generality, in this Letter we restrict our studies of biosensing performance for the laser-printed porous microrings by measuring their Raman response from a self-organized monolayer of R6G molecules and mapping surface intensity distribution of their specific main bands. A commercial Raman microscope (Alpha, WiTec), utilizing a 532-nm CW semiconductor laser source focused by a 0.9-NA dry lens (100x, Carl Zeiss) and a grating-type spectrometer (600 lines/mm) with a electrically cooled CCD camera, was used to measure SERS performance of the fabricated microrings. An ethanol solution of R6G with a concentration of 10$^{-5}$M was drop-casted on the sample surface and after its complete drying was rinsed in a distilled water producing a monolayer. All Raman spectra were measured at the 1-mW excitation laser power and the accumulation time of 0.5 s per point. Similar parameters were used for mapping of the surface intensity distribution of the main R6G Raman bands near the several smooth and porous microrings within the 2x2 $\mu$m$^2$ square sample area with 0.2-$\mu$m sampling step. The build-in software (WiTec Project) was used to obtain the 2D maps related to each spectrally narrow Raman band.

Typical Raman spectra acquired from the R6G layer on both smooth and porous microrings, having almost the same outer diameter and the wall thickness, reveal all main Raman bands characteristic for R6G molecules, while these ``fingerprints" can not be identified on the glass substrate even under 10 times increased accumulation time (Fig.3(a)). This measurement indicates strong SERS performance of the fabricated structures. The maximal Raman intensity averaged over all main spectral bands marked in Fig.3(a) for smooth microrings reaches 300$\pm 60$ counts(s$^{-1}$mW$^{-1}$), while increases to 1550$\pm 170$ counts(s$^{-1}$mW$^{-1})$ for porous microrings, yielding in the average SERS enhancement factor of $\approx 10^5$ when weighted as signal per molecule of R6G solution \cite{Kuchmizhak16}. Also, it points out a clear plasmonic contribution to the SERS enhancement, rather than the concentration effect. Mapping the surface distribution of the Raman intensity at the 1361-cm$^{-1}$ band of the R6G molecules, covering the smooth microring, reveals the position of the local electromagnetic ``hot spots'', which were found to depend on the polarization direction of the exciting laser source (orange arrows in Fig.3(b-e)).

According to our supporting FDTD calculations (for details see \cite{Kuchmizhak16}), such 532-nm, normal-incident linearly-polarized laser irradiation provides low-efficiency excitation of the smooth microring walls with quite moderate 5-fold enhancement of the squared electric field amplitude (Fig.3(d)), while the corresponding hot spots, following the linear polarization direction, are distributed non-uniformly along the ring circumference, yielding in similar distribution of SERS signal (Fig.3(b)). On the contrary, according to our FDTD calculation performed for the porous microring, nanosized surface features produce multiple strongly enhanced E-fields, homogeneously distributed along the microring surface independently on the polarization direction (Fig.3(e)) and generally correlating with the acquired SERS map (Fig.3(c)).

In conclusion, in this work we have fabricated porous microrings from thin gold films deposited in nitrogen and nitrogen-argon mixed atmospheres, by their ns-laser ablation and the following ion-beam etching post-procedure, and studied their SERS performance. The proposed approach allows us to fabricate both smooth and porous microrings with variable geometric parameters as well as tunable inner structure (nanopore size and density). The developed laser-printing/ion-beam etching procedure is demonstrated to be a versatile tool for fabricating multifunctional plasmonic structures and metasurfaces for spectroscopic bioidentification based on combination of surface-enhanced infrared absorption, SERS and photoluminescence detection schemes on a single sensing substrate.

The authors acknowledge the financial support from RFBR (A.N. --  grant No. 16-32-00365, A.K.-- grant No. 17-02-00571), the RF President grant (A.K. -- contract No. 3287.2017.2), the ``Far East Program", the RF Ministry (E.P. -- project \#3.7383.2017.8.9), and RF Government (S.I.K. -- Grant 074-U01).

\bibliography{References}

\begin{thebibliography}{48}
\bibitem [{not()}]{note2}%
  \BibitemOpen
  \href@noop {} {\bibinfo  {journal} {Deposition of all films was performed at
  a fixed applied voltage of 2.5 kV, while maintaining the current at a
  constant value of 25 mA by dynamically adjusting the discharge gas pressure}\
  }\BibitemShut {NoStop}%
\bibitem [{not({\natexlab{a}})}]{note3}%
  \BibitemOpen
\bibfield  {journal} {  }\href@noop {} {\bibfield  {journal} {\bibinfo
  {journal} {Appearance of the huge pores with the diameter larger than 150 nm
  for the structures produced in the ``nitrogen-deposited" films (Fig.2(b),
  (bottom)) can be potentially explained by merging of neighboring nanopores
  during the boiling process.}\ } ({\natexlab{a}})}\BibitemShut {NoStop}%
\bibitem [{not({\natexlab{b}})}]{note4}%
  \BibitemOpen
  \href@noop {} {\bibfield  {journal} {\bibinfo  {journal} {The non-zero
  nitrogen level detected for``Ar-deposited" films can be attributed to both
  certain residual air (nitrogen+oxygen) content during our low-vacuum
  sputtering with Ar discharge gas and nitrogen adsorption/absorption during
  storage in ambient environment}\ } ({\natexlab{b}})}\BibitemShut {NoStop}%
\end{thebibliography}

\end{document}